\documentclass{sig-alternate}
\usepackage{blindtext}
\usepackage{hyperref}
\usepackage{color,soul}
\usepackage{amsmath}
\usepackage{tcolorbox}
\usepackage{amsmath}
\usepackage{algorithm}
\usepackage[noend]{algpseudocode}
\usepackage{multirow}
\usepackage{graphicx} 

\DeclareGraphicsExtensions{.pdf,.png,.jpg}

\begin{document}
%


\CopyrightYear{2017} 

\title{An Artificial Neural Network-based Stock Trading System Using Technical Analysis and Big Data Framework}

%
%
%
%
%

\numberofauthors{3} 
%
\author{
%
%
\alignauthor
Omer Berat Sezer\\
       \affaddr{TOBB University of Economics and Technology}\\
      \affaddr{Ankara, Turkey}\\
       \email{oberatsezer@etu.edu.tr}
\alignauthor
 A. Murat Ozbayoglu\\
       \affaddr{TOBB University of Economics and Technology}\\
       \affaddr{Ankara, Turkey}\\
       \email{mozbayoglu@etu.edu.tr}
\alignauthor Erdogan Dogdu\\
       \affaddr{Georgia State University (adj.)}\\
       \affaddr{Cankaya University}\\
       \affaddr{Ankara, Turkey}\\
       \email{edogdu@cankaya.edu.tr}
}


\maketitle
\begin{abstract}
In this paper, a neural network-based stock price prediction and trading system using technical analysis indicators is presented. The model developed first converts the financial time series data into a series of buy-sell-hold trigger signals using the most commonly preferred technical analysis indicators. Then, a Multilayer Perceptron (MLP) artificial neural network (ANN) model is trained in the learning stage on the daily stock prices between 1997 and 2007 for all of the Dow30 stocks. Apache Spark big data framework is used in the training stage. The trained model is then tested with data from 2007 to 2017. The results indicate that by choosing the most appropriate technical indicators, the neural network model can achieve comparable results against the Buy and Hold strategy in most of the cases. Furthermore, fine tuning the technical indicators and/or optimization strategy can enhance the overall trading performance.
\end{abstract}


\terms{Algorithmic trading, trading strategy, machine learning, neural networks, stock market technical analysis}


\section{Introduction}
\label{sec:introduction}

Stock market forecasting and developing profitable trading models have always attracted researchers and practitioners \cite{atsalakis2009surveying}. However, it is very challenging to come up with a model that works reliably under different market conditions. In the last few decades, thanks to the advancements in computer and communications technologies, computational intelligence models started emerging as viable alternatives to the traditional decision support systems. Previous models are mostly based on static rules and analyses, hence can easily be outdated. At the same time, due to the excessive manual interactions, these models are not immune from human emotions, resulting in inconsistent, poor returns. Computational intelligence models on the other hand, such as neural networks~\cite{wang2011forecasting,liao2010forecasting,park2013stock}, neuro-fuzzy models~\cite{zarandi2009forecasting}, support vector machines (SVM)~\cite{kim2013financial}, and genetic algorithms-based systems~\cite{arajuo2013a}, demonstrated good performance achievements. Creamer utilized machine-learning algorithms to develop effective trading strategies and accordingly built automated trading agents. The agent they developed generated higher profits using Logitboost method than the simple ``buy and hold" strategy~\cite{creamer2012model}. But their experiments are very limited. They tested on two European index futures (FESX and FDAX) for only 21 trading days for March of 2009.

As implementing machine learning models using big data is becoming mainstream, 
such models started to emerge as part of algorithmic trading systems that now originates the majority of all transactions 
executed in NYSE. In this paper we aim to create such a profitable model using technical analysis indicators as features for a neural network model.
Section 2 explains the model and the features we use, section 3 presents the method and the results are evaluated in section 4.

\section{MODEL FEATURES}
\label{sec:related_work}

Technical analysis indicators have been used for identifying appropriate entry-exit points for trading models. Even though there are over 100 different indicators, some of them are more frequently used then the others, mostly because of easiness and/or effectiveness. Three of the most commonly used technical indicators are RSI, MACD and Williams \%R. These are the features that we selected to use in our neuro-trading model due to their wide acceptance. Below we briefly explain these indicators.

\subsection{Relative Strength Index (RSI)}
\label{subsec:rsi}
Relative Strength Index (RSI) is a technical momentum indicator that shows historical strength or weakness of stock prices. It also compares losses and gains in a specified time period as follows. 

\begin{equation}\label{eq:rsi} 
RSI=100 – \frac{100}{(1 + RS)}
\end{equation}

\begin{equation}\label{eq:rs} 
RS= \frac{Average Gain}{Average Loss}
\end{equation}

\subsection{Moving Average Convergence and Divergence (MACD)}
\label{subsec:macd}
MACD is a technical indicator that illustrates the trend of the stock prices. It is equal to the difference of the 12-day Exponential Moving Average (EMA) and 26-day EMA. 
\begin{equation}\label{eq:macd} 
MACD= (12DaysEMA - 26DaysEMA)
\end{equation}

\subsection{Williams \%R}
\label{subsec:williams}
Williams \%R is momentum based technical indicator that shows the overbought and oversold conditions for stock prices.
\begin{equation}\label{eq:williams} 
\%R = \frac{(Highest High - CurrentClose)} { (Highest High - Lowest Low)} x -100
\end{equation}

\section{METHOD}
\label{sec:work_description}

\begin{figure}
\centering
\includegraphics[width=0.35\textwidth]{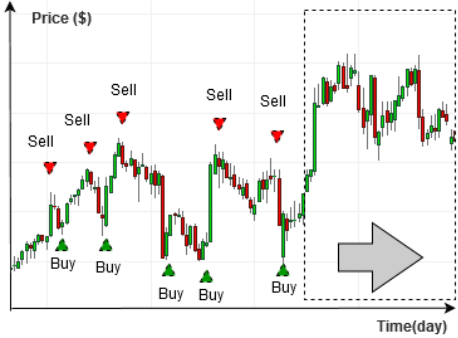}
\caption{Labelling Points with Window}
\label{fig_labelling}
\end{figure}

For big data analytics, commonly used open source tools are Apache Hadoop\footnote{http://hadoop.apache.org} and Apache Spark\footnote{http://spark.apache.org}. In our work, we used Spark. Apache Spark has a built-in machine learning library called MLlib implementing many algorithms. In general, for analyzing time series data and forecasting problems, Recurrent Neural Networks (RNN) are used in the literature. Meanwhile, Multilayer Perceptron (MLP) is also used for time series forecasting when appropriate feature processing is implemented. For this purpose, in our study, we used the aforementioned technical indicator outcomes as model features.  Furthermore, we used Spark MLlib library's MLP classifier to analyse time-series data.

In our approach, we aim to predict buy and sell (entry-exit) points of the stock prices by using MLP artificial neural network.  There are three main phases in our model for predicting the buy-sell points from the stock prices. There is an extra Phase 4 for calculating the efficiency of the system, however, that is not part of the trading model. Algorithm~\ref{algorithm} shows the steps of all phases.

\begin{figure*}[ht]
\centering
\includegraphics[width=0.48\textwidth]{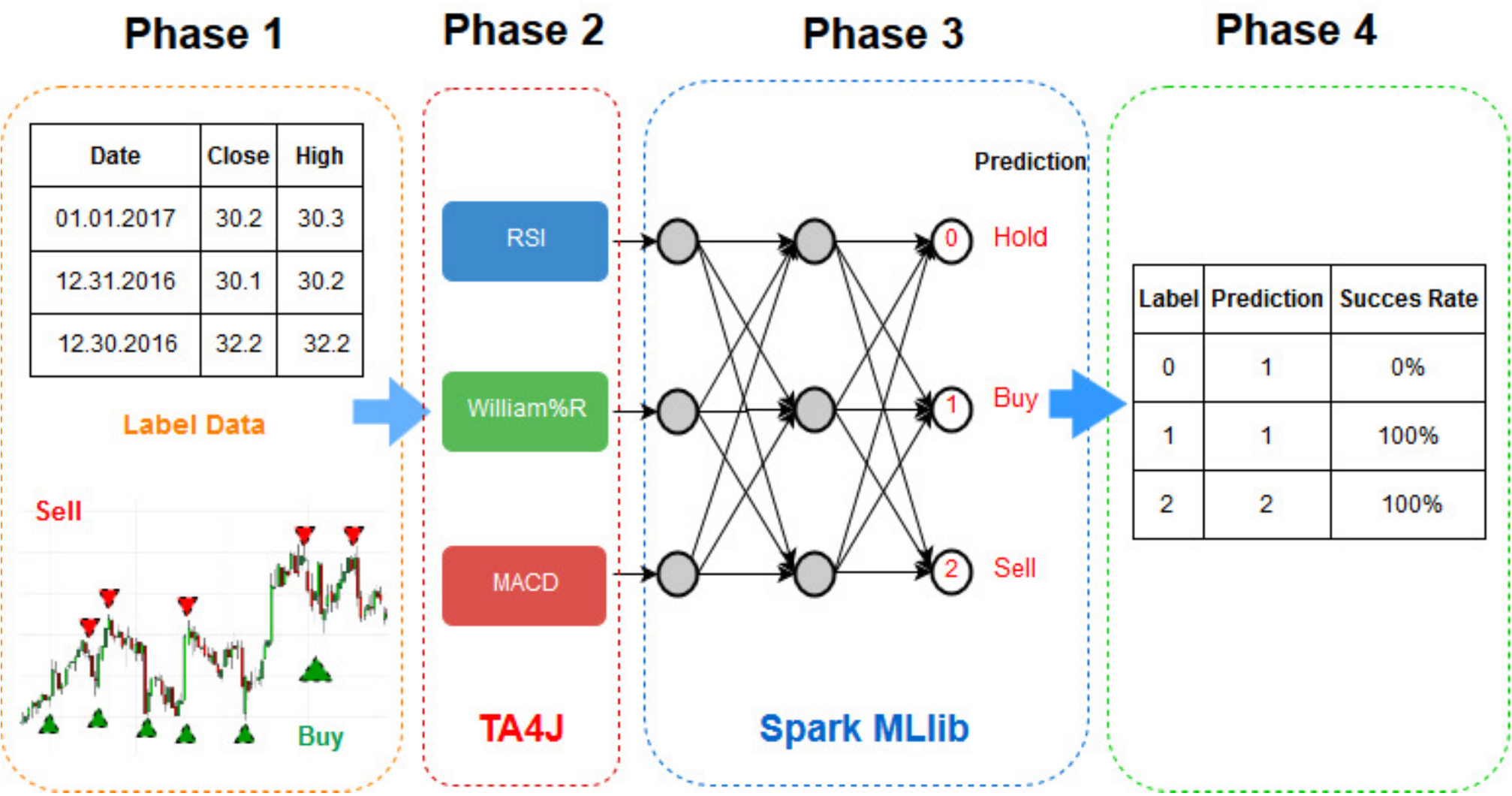}
\caption{Phases of Our Algorithm}
\label{fig_phases}
\end{figure*}

In our study, the daily stock prices for Dow 30 stocks, which are obtained from finance.yahoo.com, are used as training and test datasets. In the first phase, open and close prices, daily high and low price values for each stock are obtained. Then, for each day, all daily close prices are labelled as ``Hold", ``Buy", or ``Sell" by automatically analyzing the peak and valley points indicating highest and lowest points for a specified period. The peak points are marked as ``Sell", the valley points are marked as ``Buy" and the remaining points are marked as ``Hold" (Figure~\ref{fig_labelling})\footnote{Base graph is adapted from https://investing.com}. 

In the second phase,  RSI, WilliamR and MACD values are calculated for each daily stock price. In our framework, we used TA4J\footnote{https://github.com/mdeverdelhan/ta4j} (Techical Analysis For Java) library to calculate the RSI, WilliamR and MACD of the daily prices. Afterwards, corresponding label value, close price, RSI, WilliamR and MACD values are normalized in order to be suitable for the learning stage. 


\begin{algorithm*}[ht]
\centering
\caption{Predicting label of Dow30 Stocks using MLP}
\label{algorithm}
\begin{algorithmic}[1]
\Procedure{AllPhases()}{}
\State $Phase~1:$
\State $dataset = read (open, close, high, low, adjustedClose, volume)$
\State $dataset.adjustRatio = dataset.close / dataset.adjustedClose$
\State $adjust (dataset.open, dataset.close, dataset.high, dataset.low)~with ~adjustRatio$
\State $calculate~Label~(Buy/Sell/Hold)$ 
\State $Phase~2: $
\State $calculate~RSI, WilliamR, MACD~for~each~line~in~dataset$
\State $trainingDataset, testDataset = dataset.split(dates=1997-2006, dates=2007-2016)$
\State $trainingDataset = resample(trainingDataset)$
\State $Phase~3:$
\State $model = MLP(layers=[4,5,4,3], epochs=200, blocksize=128, seed=1234L)$
\State $model.train(trainingDataset)$
\State $model.test(testDataset)$
\State $Phase~4:$
\State $evaluateResults()$
\EndProcedure
\end{algorithmic}
\end{algorithm*}

In addition, in the second phase, the data imbalance problem is also solved. Normally, the occurrence of ``Hold" labels is much greater than the number of the ``Sell" and ``Buy" labels in the training data. This effects the learning stage such that the model only learns the majority classes better (Hold), ignoring the smaller classes causing misclassification and misprediction of data. There are different solutions in literature for this problem. We preferred the approach of resampling the minority classes. In other words, we created multiple copies of ``Buy" and ``Sell" labeled data and introduced those to the training dataset. Thus, the number of three class labels are approximately equal solving the class imbalance problem. 

In the third phase, training and test data are fed to the multilayer perceptron (MLP) using Apache Spark. Our topology for MLP has four layers that consist of 4 nodes in the input layer, 5 nodes in the second layer, 4 nodes in the third layer and 3 nodes in the fourth layer (one for each output class Buy, Hold, and Sell). MLP is run with 200 epochs to train the learning model using the labeled training data. Figure~\ref{fig_phases} depicts the whole process graphically.


\section{Evaluation}
\label{sec:evaluation}

As explained in Section~\ref{sec:work_description}, we used the three phases of the algorithm to train and test the model, and then measure the overall performance in Phase 4. The stock data obtained for Dow 30 is split into two sets, the training data is the stock prices between the dates of 1/1/1997 and 12/31/2006, and the test data is the stock prices from 1/1/2007 to 1/1/2017. We analyzed the performance of our framework with different criteria. Our focus is on the correct prediction of labels as an output of MLP model. 
Walmart (WMT) stock is chosen to provide an example for the evaluation. Table~\ref{table_confusion} illustrates the confusion matrix for WMT, and Table~\ref{table_evaluation} shows the precision, recall and F1 scores. The overall prediction rate accuracy for WMT is 65.52\%.


\begin{table}[]
\centering
\caption{Confusion Matrix of WMT (Walmart Stock)}
\label{table_confusion}
\begin{tabular}{|c|l|l|l|l|}
\hline
\multicolumn{1}{|l|}{}           & \multicolumn{4}{c|}{\textbf{Predicted}}           \\ \hline
\multicolumn{1}{|l|}{}           &            & \textbf{0} & \textbf{1} & \textbf{2} \\ \hline
\multirow{3}{*}{\textbf{Actual}} & \textbf{0} & 889        & 429        & 868        \\ \cline{2-5} 
                                 & \textbf{1} & 41         & 110        & 4          \\ \cline{2-5} 
                                 & \textbf{2} & 21         & 0          & 139        \\ \hline
\end{tabular}
\end{table}

\begin{table}[]
\centering
\caption{Evaluation Of WMT}
\label{table_evaluation}
\begin{tabular}{|l|l|l|l|}
\hline
                   & \textbf{Class 0} & \textbf{Class 1} & \textbf{Class 2} \\ \hline
\textbf{Precision} & 0.93             & 0.20             & 0.14             \\ \hline
\textbf{Recall}    & 0.41             & 0.71             & 0.87             \\ \hline
\textbf{F1 Score}  & 0.57             & 0.32             & 0.24             \\ \hline
\end{tabular}
\end{table}
Moreover, we evaluated our system based on the success of our trading strategy. In our model, a stock is bought, sold or held according to its predicted label result. For instance, if the predicted label equals to "1" (buy), (the corresponding output neuron is activated) the stock is bought using the capital that exists at that particular point. We start with a total capital of \$10,000. All available capital is used during each transaction. If the predicted label equals to "2" (sell), the stock is sold and we get back to an all cash position. If predicted label equals to "0" (hold), system does not do anything. As a result, during trading, if the same label is repeated one after another, only the first label gets triggered, the system ignores the repeating signals until the label changes. Also, in our scenario, we used a realistic trading environment that includes trading commission (\$1 per transaction, 0.001 of the starting capital). Stop loss situations (\%5) are also implemented in our scenario. 

\begin{tcolorbox}
\small{
Transaction Number, Interval, Gain, Instant Capital
\\1.(21-25) => -516.19 Capital: \$9481.81
\\..
\\62.(831-836) => -1532.45 Capital: \$19428.23
\\..
\\168.(2463-2465) => 1061.5 Capital: \$49181.78
}
\end{tcolorbox}

The box above shows a sample of JPM (JPMorgan) trades by our framework. There are 168 transactions for JMP between 2007-2017. Starting capital is \$10,000 in 1/1/2007. The ending capital reaches \$49,181.78 at the end.

We also applied ``Buy and Hold" (BaH) as the base strategy for Dow 30 stocks for the same testing period.  In BaH, a stock is bought at the beginning and sold at the end of the testing period. This is the preferred strategy for most long-term investors and works very well for bull markets, but not so good in trendless or bear markets.

\begin{table*}[!htb]
\centering
\caption{Evaluation of Our Proposed System with Dow30 Shares - Total Capital with Our Prosed Strategy (OUR), Total Capital with Buy and Hold Strategy (BaH), Our Annualized Return (OURr), Buy and Hold Annualized Return (BaHr), Annualized Number of Transaction (AnT), Percent of Success (PoS), Average Percent Profit Per Transactions (ApT), Average Transaction Length (L), Maximum Profit Percentage in Transaction (MpT), Maximum Loss Percentage in Transaction (MlT), Maximum Capital (MxC)}
\label{table:dow30}
\begin{tabular}{l r r r r r r r r r r r }
\hline
\textbf{Share} & \textbf{OUR} & \textbf{BaH} & \textbf{OURr} & \textbf{BaHr} & \textbf{AnT} & \textbf{Pos} & \textbf{ApT} & \textbf{L} & \textbf{MpT} & \textbf{MlT} & \textbf{MxC} \\ \hline
\textbf{MMM}   & \$15234.16   & \$29324.88   & 6.33\%        & 16.99\%       & 12.0           & 67.07\%      & 0.63\%       & 5          & 12.54\%      & -8.20\%      & \$15505.23   \\ \hline
\textbf{AXP}   & \$14727.15   & \$15157.78   & 5.80\%        & 6.25\%        & 15.6         & 57.01\%      & 0.68\%       & 13         & 43.09\%      & -13.75\%     & \$21180.43   \\ \hline
\textbf{AAPL}  & \$14742.93   & \$104256.20   & 5.83\%        & 40.79\%       & 5.8          & 60.00\%      & 1.26\%       & 20         & 23.28\%      & -7.33\%      & \$17804.25   \\ \hline
\textbf{BA}    & \$17010.05   & \$22809.31   & 8.05\%        & 12.78\%       & 20.3         & 66.91\%      & 0.50\%       & 5          & 6.17\%       & -8.90\%      & \$18302.59   \\ \hline
\textbf{CAT}   & \$10252.42   & \$21030.51   & 0.36\%        & 11.44\%       & 31.0           & 62.91\%      & 0.12\%       & 4          & 11.54\%      & -8.31\%      & \$12895.92   \\ \hline
\textbf{CVX}   & \$17907.21   & \$22968.13   & 8.87\%        & 12.89\%       & 20.6         & 67.38\%      & 0.48\%       & 3          & 12.53\%      & -9.98\%      & \$18349.65   \\ \hline
\textbf{CSCO}  & \$21182.93   & \$13126.52   & 11.57\%       & 4.05\%        & 22.0           & 66.89\%      & 0.64\%       & 7          & 8.96\%       & -9.39\%      & \$21594.89   \\ \hline
\textbf{KO}    & \$17258.98   & \$23354.41   & 8.29\%        & 13.18\%       & 8.6          & 76.27\%      & 1.03\%       & 9          & 4.53\%       & -12.17\%     & \$17258.98   \\ \hline
\textbf{DIS}   & \$28859.03   & \$34368.91   & 16.71\%       & 19.72\%       & 22.3         & 70.59\%      & 0.82\%       & 8          & 11.85\%      & -5.41\%      & \$30457.32   \\ \hline
\textbf{DD}    & \$17750.91   & \$22197.39   & 8.74\%        & 12.34\%       & 18.0           & 66.67\%      & 0.60\%       & 5          & 19.97\%      & -6.65\%      & \$19297.54   \\ \hline
\textbf{XOM}   & \$18385.49   & \$15946.05   & 9.30\%        & 7.05\%        & 23.7         & 66.67\%      & 0.47\%       & 5          & 20.27\%      & -5.78\%      & \$18868.03   \\ \hline
\textbf{GE}    & \$12663.52   & \$12399.64   & 3.50\%        & 3.19\%        & 21.9         & 65.33\%      & 0.31\%       & 6          & 15.30\%      & -14.68\%     & \$13237.47   \\ \hline
\textbf{GS}    & \$14230.22   & \$12238.97   & 5.28\%        & 2.99\%        & 23.2         & 64.78\%      & 0.39\%       & 5          & 24.93\%      & -15.50\%     & \$14230.22   \\ \hline
\textbf{HD}    & \$15088.71   & \$43768.70    & 6.19\%        & 24.04\%       & 24.7         & 68.64\%      & 0.36\%       & 6          & 2.89\%       & -7.65\%      & \$18299.88   \\ \hline
\textbf{IBM}   & \$17151.82   & \$21143.52   & 8.19\%        & 11.55\%       & 17.2         & 70.34\%      & 0.54\%       & 5          & 5.66\%       & -7.86\%      & \$19265.33   \\ \hline
\textbf{INTC}  & \$27965.75   & \$23656.29   & 16.21\%       & 13.40\%       & 22.2         & 68.42\%      & 0.80\%       & 6          & 7.05\%       & -7.18\%      & \$31877.34   \\ \hline
\textbf{JNJ}   & \$19043.10    & \$23687.77   & 9.86\%        & 13.42\%       & 17.8         & 73.77\%      & 0.58\%       & 6          & 8.85\%       & -6.99\%      & \$19279.80    \\ \hline
\textbf{JPM}   & \$49181.78   & \$22092.57   & 26.17\%       & 12.26\%       & 24.5         & 67.26\%      & 1.21\%       & 5          & 27.14\%      & -8.62\%      & \$49181.78   \\ \hline
\textbf{MCD}   & \$17519.35   & \$38489.77   & 8.53\%        & 21.75\%       & 15.5         & 70.75\%      & 0.56\%       & 2          & 3.79\%       & -4.01\%      & \$18445.34   \\ \hline
\textbf{MRK}   & \$29081.32   & \$18865.70    & 16.86\%       & 9.71\%        & 22.5         & 69.48\%      & 0.79\%       & 6          & 8.86\%       & -6.71\%      & \$29389.65   \\ \hline
\textbf{MSFT}  & \$37923.78   & \$25820.00    & 21.48\%       & 14.85\%       & 22.6         & 69.03\%      & 0.97\%       & 6          & 6.28\%       & -5.77\%      & \$37923.78   \\ \hline
\textbf{NKE}   & \$22940.48   & \$48496.06   & 12.89\%       & 25.93\%       & 16.9         & 67.24\%      & 0.93\%       & 13         & 28.39\%      & -8.50\%      & \$28257.09   \\ \hline
\textbf{PFE}   & \$11094.86   & \$18953.47   & 1.53\%        & 9.78\%        & 22.9         & 64.33\%      & 0.16\%       & 6          & 7.07\%       & -8.52\%      & \$11653.85   \\ \hline
\textbf{PG}    & \$20278.23   & \$17434.55   & 10.88\%       & 8.46\%        & 18.8         & 68.99\%      & 0.62\%       & 9          & 10.23\%      & -5.48\%      & \$20278.23   \\ \hline
\textbf{TRV}   & \$64371.78   & \$31098.53   & 31.23\%       & 18.01\%       & 24.1         & 75.15\%      & 1.26\%       & 7          & 33.97\%      & -6.54\%      & \$64371.78   \\ \hline
\textbf{UTX}   & \$18540.16   & \$20932.55   & 9.43\%        & 11.38\%       & 24.1         & 68.48\%      & 0.47\%       & 5          & 21.00\%      & -10.23\%     & \$19360.46   \\ \hline
\textbf{UNH}   & \$9343.90     & \$34464.65   & -0.99\%       & 19.80\%       & 15.2         & 57.69\%      & 0.25\%       & 9          & 10.67\%      & -17.73\%     & \$12030.98   \\ \hline
\textbf{VZ}    & \$12147.37   & \$24315.17   & 2.88\%        & 13.83\%       & 16.2         & 61.26\%      & 0.26\%       & 6          & 22.05\%      & -5.97\%      & \$13267.72   \\ \hline
\textbf{WMT}   & \$32230.01   & \$18389.92   & 18.63\%       & 9.30\%        & 18.5         & 73.23\%      & 0.98\%       & 7          & 11.07\%      & -8.07\%      & \$32230.01   \\ \hline
\end{tabular}
\end{table*}

Table~\ref{table:dow30} shows the comparative performance of our model against the BaH for all Dow 30 stocks.
Our proposed framework's average annualized return is 10.3\%, and the average annualized return of BaH strategy is 13.83\%. Our proposed strategy's annualized return performed better than BaH strategy's annualized return in only 9 out of 29 (Visa stock [V] did not have enough data points in the same period). The average success percentage of all transactions (buy then sell) in our system is 67.33\% indicating that every 2 out of 3 transactions resulted in a profit.

Generally it is very difficult to beat Buy and Hold during such a lengthy time period. However, our model provided mixed results (sometimes better, sometimes worse) in comparison to BaH. This is mainly because of using the same standard values for the chosen technical parameters for all of the stocks. Since, we have not implemented any new indicator and/or parameter optimization, the performances of the stocks vary accordingly. However, it is seen in previous studies that implementing such optimization techniques might increase the overall trading performance considerably~\cite{ozbayoglu2010stock}. Fine tuning the technical indicator parameters individually for each stock might improve the overall performance of the trading model.  

\section{Conclusion}
\label{sec:conclusion}

In this paper, we presented a new stock trading and  prediction model based on a MLP neural network utilizing technical analysis indicator values as features. Big data framework Apache Spark is used in implementation. The model is trained and tested on Dow 30 stocks in order to see the evaluate the model. The results indicate that comparable results are obtained against the baseline Buy and Hold strategy even without fine tuning and/or optimizing the model parameters. For future work, some optimization stages will be added to the model and deep learning models will be used in the learning stage.

%
\bibliographystyle{abbrv}
\bibliography{sigproc}  
%
%

\end{document}